\documentclass[prb,aps,twocolumn,superscriptaddress]{revtex4}
\usepackage{amsmath} 
\usepackage{amssymb} 
\usepackage{amsfonts}
\usepackage[]{epsfig} 
\begin{document}
\bibliographystyle{apsrev} 

\title{Interface dynamics of microscopic cavities in water} 
\author{Joachim Dzubiella}
\email[e-mail address:] {jdzubiel@ph.tum.de} 
\affiliation{Physics Department, Technical University  Munich, 85748 Garching, Germany} 
\date{\today}

\begin{abstract}
  An analytical description of the interface motion of a collapsing
  nanometer-sized spherical cavity in water is presented by a
  modification of the Rayleigh-Plesset equation in conjunction with
  explicit solvent molecular dynamics simulations. Quantitative
  agreement is found between the two approaches for the time-dependent
  cavity radius $R(t)$ at different solvent conditions while in the
  continuum picture the solvent viscosity has to be corrected for
  curvature effects. The typical magnitude of the interface or
  collapse velocity is found to be given by the ratio of surface
  tension and fluid viscosity, $v\simeq\gamma/\eta$, while the
  curvature correction accelerates collapse dynamics on length scales
  below the equilibrium crossover scales ($\sim$1nm). The study offers
  a starting point for an efficient implicit modeling of water
  dynamics in aqueous nanoassembly and protein systems in
  nonequilibrium.
\end{abstract}

\pacs{68.03.-g,68.35.Md,83.10.Rs,47.55.dd}

\maketitle

\section{Introduction}

Hydrophobic hydration in equilibrium is a phenomenon which exhibits
qualitatively different behavior at small and large length scales.
\cite{chandler:review,rajamani} While small solutes (radii
$R\lesssim$1nm) are accommodated by water with only minor
perturbations, larger solutes ($R\gtrsim$1nm) induce major
rearrangements of water interfacial structure. As a consequence the
solvation free energy $G(R)$ of small hydrophobic cavities scales with
solute volume while for larger cavities it grows with surface area (as
a good approximation near liquid-vapor coexistence) accompanied by
weak solvent dewetting at extended restraining hydrophobic
surfaces.\cite{poynor} Furthermore, water, which is close to the
liquid-vapor transition at normal conditions, can minimize interface
area by locally evaporating and forming a 'nanobubble' within
hydrophobic confinement. Evidence of bubble formation in confined
geometry has been given early by computer simulations of smooth
plate-like solutes,\cite{wallquist:jpc} but more recently it has been
demonstrated in varying degrees in atomistically resolved plate-like
solutes, \cite{koishi, rossky:pre} hydrophobic tubes and ion
channels,\cite{beckstein:proc,sukharev} and in the collapse of
proteins,\cite{zhou:science,berne:nature} suggesting that it plays a
key role in the stabilization and folding dynamics of certain classes
of biomolecules.\cite{tenwolde, huang2:pnas} Experimental evidence of
nanobubbles in strong confinement (in contrast to bubbles at a single
planar surface~\cite{poynor}) has been given for instance in studies
of water between hydrophobic surfaces,\cite{attard:prl} in zeolites
and silica nanotubes,\cite{helmy,jayaraman} and on a subnanometer
scale in nonpolar protein cavities.\cite{collins}

The dewetting induced change in solvation energy is typically
estimated using simple macroscopic arguments as known from capillarity
theory, e.g. by describing interfaces with Laplace-Young (LY) type of
equations.\cite{huang:pnas,helmy} Recently an extension of the LY
equation has become available which extrapolates to microscopic scales
by including a curvature correction to the interface tension and
considering atomistic dispersion and electrostatic potentials of the
solvated solute explicitly.\cite{dzubiella:prl} Although those
macroscopic considerations (e.g,. the concept of surface tension) are
supposed to break down on atomistic scales they show surprisingly good
results for the solvation energy of microscopic solutes, e.g.~alkanes
and noble gases, and quantitatively account for dewetting effects in
nanometer-sized hydrophobic confinement.\cite{dzubiella:jcp} While we
conclude that the equilibrium location of the solute-solvent interface
seems to be well described by those techniques, nothing is known about
the interface {\it dynamics} of evolution and relaxation. In this
study we address two fundamental questions: First, what are the
equations which govern the interface motion on atomistic ($\sim$1nm)
scales?  Secondly, does the dynamics exhibit any signatures of the
length scale crossover found in equilibrium?

On macroscopic scales the collapse dynamics of a (vapor or gas) bubble
is related to the well-known phenomenon of
sonoluminescence.\cite{brennen} The governing equations can be derived
from Navier-Stokes and capillarity theory and are expressed by the
Rayleigh-Plesset (RP) equation.\cite{plesset} We will show that the RP
equation simplifies in the limit of microscopic cavities and can be
extended to give a quantitative description of cavity interface
dynamics on nanometer length scales. We find a qualitatively different
dynamics than the typical ``mean-curvature flow'' description of
moving interfaces, \cite{spohn} in particular a typical magnitude of
interface or collapse velocity given by the ratio of surface tension
and fluid viscosity, $v\simeq \gamma/\eta$. Our study is restricted to
the generic case of the collapse of a spherical cavity and is
complemented by explicit solvent molecular dynamics (MD) computer
simulations.  We note here that recently, Lugli and Zerbetto studied
nanobubble collapse in ionic solutions by MD simulations on similar
length scales.\cite{lugli} While their MD data compares favorably with
our results their interpretation and conclusions in terms of the RP
equation are different.  We will resume this discussion in the
conclusion section. 

In this study we show that a simple analytical approach quantitatively
describes microscopic cavity collapse for a variety of different
solvent situations while the simulations suggest that the solvent
viscosity needs to be corrected for curvature effects. Our study might
offer a simple starting point for an efficient implicit modeling of
water dynamics in aqueous nanoassembly and protein systems in
nonequilibrium.

\section{Theory}

The Rayleigh-Plesset equation for the time evolution of a macroscopic
vapor bubble with radius $R(t)$ can be written as\cite{plesset}
\begin{eqnarray}
-\rho_m\left(R\ddot R+\frac{3}{2}\dot R^2\right)=\Delta P+4\eta\frac{\dot R}{R}+\frac{2\gamma}{R},
\label{RP1}
\end{eqnarray}
where $\rho_m$ is the solvent mass density, $\Delta P=P-P_{\rm v}$ the
difference in liquid and vapor pressures, $\eta$ the dynamic
viscosity, and $\gamma$ the liquid-vapor interface tension. While for
macroscopic bubble radii the inertial terms (left hand side) control
the dynamics, for decreasing radii the frictional and pressure terms
(right hand side) grow in relative magnitude and eventually dominate,
so that completely overdamped dynamics can be assumed on atomistic
scales:
\begin{eqnarray}
\dot R \simeq -\frac{R}{4\eta}\left(\Delta P +\frac{2\gamma}{R}\right).
\label{RP2}
\end{eqnarray}
A rough estimate for the threshold radius $R_t$ below which friction
dominates is given when the Reynolds number ${\cal R}=vR\rho_m/\eta$
becomes unity and viscous and inertial forces are balanced. With a
typical initial interface velocity of the order of $v\sim\gamma/\eta$
[from $\ddot R(0)= 0$ in eq.~(\ref{RP1})] we obtain
\begin{eqnarray}
R_{t}=\eta^2/(\rho_m \gamma)
\end{eqnarray}
which is $\simeq 10$nm for water at normal conditions.  Note that this
threshold value can deviate considerably for a fluid different than
water and that the viscosity typically has a strong temperature ($T$)
dependence which implies that $R_t$ can change significantly with $T$.

In equilibrium ($\dot R=0$) the remaining expression in
eq.~(\ref{RP2}) is the (spherical) LY equation $\Delta P+2\gamma/R=0$.
Thus eq.~(\ref{RP2}) describes a linear relationship between capillary
pressure and interface velocity where $R/(4\eta)$ plays the role of an
interface mobility (inverse friction).\cite{spohn}  Interestingly,
the mobility is linear in bubble radius which leads to a {\it
  constant} velocity driven by surface tension {\it independent} of
radius (assuming $P\simeq0$); this has to be contrasted to the
typically used capillary dynamics which is proportional to the local
mean curvature $\propto 1/R$.\cite{spohn}

Generalizations of the LY equation to small scales are available by
adding a Gaussian curvature term ($\sim1/R^2$) as shown by Boruvka and
Neumann\cite{boruvka}; that has been demonstrated to be equivalent to
a first order curvature correction in surface tension,
i.e. $\gamma(R)=\gamma_\infty(1-\delta_{\rm
  T}/R)$,\cite{dzubiella:prl} where $\delta_{\rm T}$ is the Tolman
length~\cite{tolman} and $\gamma_\infty$ the liquid-vapor surface
tension for a planar interface ($R=\infty$). The Tolman length has a
magnitude which is usually of the order of the size of a solvent
molecule. Furthermore, it has been observed experimentally that the
viscosity of strongly confined water can depend on the particular
nature of the surface/interface.\cite{klein:jpcm} We conclude that in
general one has to anticipate that - analogous to the surface tension
- the effective interface viscosity obeys a curvature correction in
the limit of small cavities due to water restructuring in the first
solvent layers at the hydrophobic interface. In the following we make
the simple first order assumption that the correction enters
eq.~(\ref{RP2}) also linear in curvature ($\sim 1/R$) yielding
\begin{eqnarray}
\dot R &=& -\frac{R}{4\eta_\infty}\left(1+\frac{\delta_{\rm vis}}{R}\right)\left(\Delta P +\frac{2\gamma_\infty}{R}\left[1-\frac{\delta_{\rm T}}{R}\right]\right)\nonumber \\
&\simeq& -\frac{1}{4\eta_\infty}\left(\Delta P R+\Delta P\delta_{\rm vis}+2\gamma_\infty+\frac{2\delta\gamma_\infty}{R}\right),
\label{RP3}
\end{eqnarray}
where the constant $\delta_{\rm vis}$ is the coefficient for the first
order curvature correction in viscosity and $\eta_\infty$ the
macroscopic bulk viscosity. Additionally, we define
$\delta=\delta_{\rm vis}-\delta_{\rm T}$ and second order terms in
curvature are neglected. We note that the choice of the $1/R$-scaling
of the viscosity curvature correction has no direct physical
justification and is arbitrary. We think however, that a curvature
correction based on an expansion in orders of  mean curvature is the 
simplest and most natural way for such a choice.

In water at normal conditions the pressure terms in (\ref{RP3}) are
negligible so that for large radii ($R \gg \delta$) the interface
velocity is constant and $R(t)=R_0-\gamma_\infty/(2\eta_\infty)
t$. This leads to a collapse velocity of about $v\simeq$0.4\AA/ps
(40m/s) which is 6$\%$ of the thermal velocity of water $v_{\rm
  th}=\sqrt{3k_BT/m}$ showing that dissipative heating of the system
is relatively weak on these scales.  A rough estimate for the
dissipation rate can be made by the released interfacial energy ${\rm
  d}G(R,t)/{\rm d}t\simeq{\rm d}(4\pi R(t)^2 \gamma_\infty)/{\rm d}t
=-4\pi\gamma_\infty^2R(t)/\eta_\infty$ yielding for instance ${\rm
  d}G(R,t=0)/{\rm d}t\simeq -35k_BT/$ps for a bubble with
$R_0=$2nm. At small radii ($R\simeq \delta$) the solution of
(\ref{RP3}) goes as $R(t)\sim
\pm\sqrt{const-(\delta\gamma_{\infty}/\eta_{\infty}) t}$ decreasing or
increasing the velocity depending on the sign of $\delta=\delta_{\rm
  vis}-\delta_{\rm T}$, i.e. the acceleration depends on the
particular sign and magnitude of the curvature corrections to surface
tension and viscosity. For large pressures and radii the first term
dominates which gives rise to an exponential decay
$R(t)\sim\exp[-\Delta P/(4\eta_{\infty})t]$.  While extending to small
scales we have assumed that the time scale of internal interface
dynamics, i.e. hydrogen bond rearrangements,\cite{kuo:science} is much
faster than the one of bubble collapse.

\section{MD simulation}

In order to quantify our analytical predictions we complement the
theory by MD simulations using explicit SPC/E water.\cite{berendsen:jpc}  The liquid-vapor surface tension of SPC/E water
has been measured and agrees with the experimental value for a wide
range of temperatures.\cite{alejandre} For $T=300$K and $P=1$bar we
have $\gamma_{\infty}=72$mN/m. The Tolman length has been estimated to
be $\delta_{\rm T}\simeq 0.9$\AA~ from equilibrium measurements of the
solvation energy of spherical cavities.\cite{huang:jpcb:2002} At the
same conditions the dynamic viscosity of SPC/E water has been found to
be $\eta_{\infty}=6.42\cdot10^{-4}$Pa$\cdot$s,\cite{hess}
$\sim$24$\%$ smaller than for real water. In experiments in nanometer
hydrophobic confinement and at interfaces however, the viscosity shows
deviations from the bulk value but remains comparable.
\cite{klein:jpcm} We proceed by treating the viscosity $\eta_\infty$
as an adjustable parameter together with its curvature correction
coefficient~$\delta_{\rm vis}$.

The MD simulations are carried out with the DLPOLY2 package~\cite{dlpoly} using an integration time step of 2fs. The simulation
box is cubic and periodic in all three dimensions with a length of
$L=(61.1\pm0.2$)\AA~ in equilibrium involving $N=6426$ solvent
molecules. Electrostatic interactions are calculated by the
smooth-particle mesh Ewald summation method. Lennard-Jones
interactions are cut-off and shifted at 9\AA.  Our investigated
systems are at first equilibrated in the $NPT$ ensemble with
application of an external spherical potential of the form $\beta
V(r)=[{\rm \AA}/(r-R_0')]^{12}$ and all molecules removed with
$r<R_0'$ since vapor can safely be neglected on these scales. This
stabilizes a well-defined spherical bubble of radius $R_0\simeq
R_0'+1$\AA. We define the cavity radius by the radial location where
the water density $\rho(r)$ drops to half of the bulk density
$\rho_0/2$. Thirty independent configurations in 20ps intervals are
stored and serve as initial configurations for the nonequilibrium
runs. We employ a Nos{\'e}-Hoover barostat and thermostat with a 0.2ps
relaxation time to maintain the solvent at a pressure $P$ and a
temperature $T$. Other choices of relaxation times in the reasonable
range between 0.1 and 0.5ps do not alter our results. In the
nonequilibrium simulations the constraining potential is switched off
and the relaxation to equilibrium is averaged over the thirty runs.

\section{Results}

\begin{table}[h]
\begin{center}
\begin{tabular}{l | c c c c c }
  system  & $P/$bar & $T$/K & $c_{\rm NaCl}$/M & $Q/e$ & $\eta_{\infty}$/($10^{-4}$Pa$\cdot$s) \\ 
\hline 
I  & $1 $ & 300  & 0    & 0  & 5.14  \\
II & $1 $ & 300  & 1.5    &  0 & 5.94 \\
III& $1 $ & 277  & 0 &  0 & 8.48     \\
IV & 2000 & 300  & 0    &  0 &  4.56 \\
V  & 1000 & 300  & 0    &  0 &  4.72 \\
VI & $1   $ & 300  & 0    & +2 & 5.14 \\
\end{tabular}
\caption{Investigated system parameters: pressure $P$, temperature
  $T$, and salt (NaCl) concentration $c$. In system VI a fixed ion
  with charge $Q=+2e$ is placed at the center of the collapsing
  bubble. The viscosity $\eta_{\infty}$ is a fit-parameter in systems
  I-V (see text).}
\label{tab}
\end{center}
\end{table}

We perform simulations of six different systems I-VI whose features
are summarized in Tab.~I and differ in thermodynamic parameters $T$
and $P$ (I, III, IV, and V) but also inclusion of dispersed salt (II),
and the influence of a charged particle in the bubble center (VI) are
considered. Note that the exact value of the crossover length scale
(however defined) can depend on the detailed thermodynamic or solvent
condition but remains close to 1nm.\cite{rajamani}

\begin{figure}[htp]
\includegraphics[width=7cm,angle=-90]{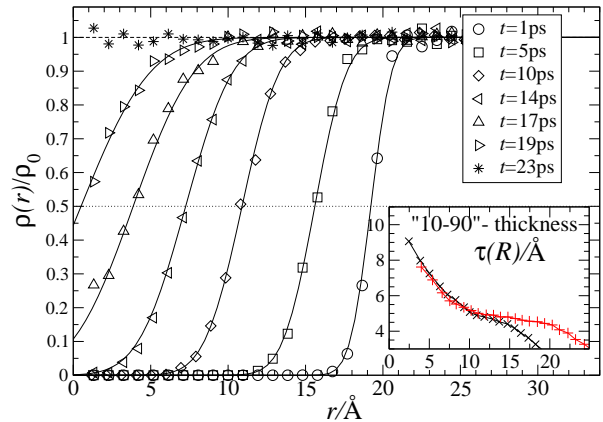}
\caption{Interface density profiles $\rho(r)/\rho_0$ for system I
  are plotted vs.~the radial distance $r$ from the bubble center for
  different times $t/$ps=1,5,10,14,17,19,23.  Symbols denote MD
  simulation data and lines are fits using $2\rho(r)/\rho_0={\rm
    erf}\{[r-R(t)]/d\}+1$. The bubble radius $R(t)$ is defined by the
  distance at which the density is $\rho_0/2$ (dotted line).  The
  inset shows the ``10-90'' thickness $\tau=1.8124\,d$ of the
  interface vs. $R$ for initial radii $R_0=19.83$\AA~ (pluses) and
  $R_0=25.6$\AA~ (crosses).}
\label{fig:1}
\end{figure}

\begin{figure}[htp]
\includegraphics[width=7cm,angle=-90]{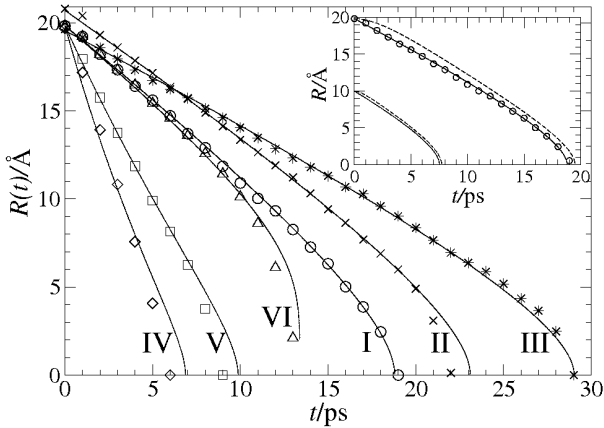}
  \caption{Time evolution of the cavity radius $R(t)$ for parameters
    as defined in systems I-VI. The solution of the modified RP
    equation (\ref{RP3}) (lines) is plotted vs. MD data (symbols).
    The inset shows the solution of the modified RP equation including
    inertia terms, cf. lhs of (\ref{RP1}), (dashed lines) compared to
    eq.~(\ref{RP3}) for system I with initial radii $R_0=19.83$\AA~
    and $R_0=10.0$\AA.}
\label{fig:2}
\end{figure}

System I is at normal conditions ($T$=300K, $P$=1bar) and consists of
pure SPC/E water. Fig.~1 shows the observed interface profiles in the
nonequilibrium situation at different times $t$/ps=1, 5, 10, 14, 17,
19, and 23 starting from an initial radius $R_0=19.83$\AA. The
liquid-vapor interface stays relatively sharp in the process of
relaxation but broadens noticeably for smaller radii. At $t\simeq
23$ps the system is completely relaxed to a homogeneous density
distribution.  The same time scale of bubble collapse has been found
in explicit water computer simulations of dewetting in nanometer-sized
paraffin plates,\cite{huang:pnas} polymers,\cite{tenwolde} and
atomistically resolved proteins.\cite{zhou:science,berne:nature}

We find that the interface profiles can be fitted very well with a
functional form $2\rho(r)/\rho_0={\rm erf}\{[r-R(t)]/d\}+1$, where $d$
is a measure of the interface thickness.  The interface fits are also
shown in Fig.~1 together with the MD data. The experimentally
accessible ``10-90'' thickness $\tau$ of an interface is the thickness
over which the density changes from $0.1\rho_0$ to $0.9\rho_0$ and is
related to the parameter $d$ via $\tau=1.8124\,d$. While experimental
values of $\tau$ for the planar water liquid-vapor interface vary
between $\sim$ 4 and 8\AA~ the measured values for SPC/E water in the
finite simulation systems are $\tau_\infty=$3 to 4\AA.\cite{alejandre}
We find a strongly radius-dependent function $\tau(R)$ plotted in the
inset to Fig.~1 for initial radii $R_0=19.83$\AA~ and $R_0=25.6$\AA.
For $R\simeq R_0$ the thickness increases during the following 5ps
from the equilibrium value $\tau\simeq 3$\AA~ to about $\tau \simeq
4.5-5$\AA~ independent of $R_0$.  While the exact equilibrium
thickness at $t=0$ depends on the particular choice of the confining
potential $V(r)$ (e.g., a softer potential might lead to a broader
initial interface) this suggest that 4.5-5\AA~ is the typical
interface thickness for a bubble of 1nm size. Regarding the slope
of the curve one might speculate that
$\tau(R\rightarrow \infty)$ saturates to the thickness $\tau_\infty$
of the measured planar interface for $R_0\rightarrow \infty$. For
$R\lesssim 10$\AA~ the thickness increases twofold during the
relaxation to equilibrium. This broadening might be attributed to
increased density fluctuations and the structural change of
interfacial water in the system when crossing from large to small
length scales which has been shown to happen in equilibrium at $\sim$
1nm.\cite{chandler:review,rajamani}

In Fig.~2 we plot the time evolution of the bubble radius $R(t)$ for
all investigated systems. Let us first focus on the simulation data of
system I (circles). As anticipated the bubble radius decreases
initially in a linear fashion while for smaller radii ($R(t)\lesssim
10$\AA) the velocity steadily increases. From the best fit of
eq.~(\ref{RP3}) we find a viscosity $\eta_{\infty}=5.14\cdot
10^{-4}$Pa$\cdot$s and its curvature correction coefficient
$\delta_{\rm vis}=4.4$\AA. Although investigating a confined system
with large interfaces the viscosity value differs only 20$\%$ from the
SPC/E bulk value. Furthermore, from our macroscopic point of view the
MD data show that high curvature decreases the viscosity and the
latter has to be curvature-corrected with a (positive) coefficient
{\it larger} than the Tolman length $\delta_{\rm T}$. If the surface
tension decreased in a stronger fashion with curvature than viscosity
the collapse velocity would drop in qualitative disagreement with the
simulation.  The overall behavior of $R(t)$ and the collapse velocity
of about $\sim$ 1\AA/ps agrees very well with the recent MD data of
Lugli and Zerbetto, who simulated the collapse of a 1nm sized bubble
in SPC water.\cite{lugli}

The inset to Fig.~2 shows the solution of eq.~(\ref{RP3}) including
inertial terms [left hand side of (\ref{RP1})] to check the assumption
of overdamped dynamics. While inertial effects are indeed small but
not completely negligible for an initial radius $R_0=19.83$\AA~ they
basically vanish for $R_0=10$\AA. Interestingly, the inertial effects
are not visible in the MD simulation data at all. We attribute this
observation to the finite and periodic simulation box which is known
to suppress long-ranged inertial (hydrodynamic)
effects.\cite{duenweg}

In the following we assume $\delta_{\rm vis}$ to be {\rm independent}
of the other parameters and treat only $\eta_{\infty}$ as adjustable
variable. In system II we add 175 salt pairs of sodium chloride (NaCl)
into the aqueous solution resulting in a concentration of $c\simeq
$1.5M. The ion-SPC/E interaction parameters are those used by Bhatt
{\it et al.}\cite{bhatt} who measured a linear increase of surface
tension with NaCl concentration in agreement with experimental
data. While this increment for $c=1.5$M is about small 2-3$\%$, the
viscosity has been measured experimentally to increase by
approximately $18\%$ at 298.15K.\cite{lang} Indeed by comparing the
simulation data to the theory we find a $16\%$ larger viscosity
$\eta_{\infty}=5.94\cdot 10^{-4}$Pa$\cdot$s. A slower collapse
velocity has been found also in the MD simulations of Lugli and
Zerbetto in concentrated LiCl and CsCL solutions when compared to pure
water.\cite{lugli}

In system III we investigate the effect of lowering the temperature by
simulating at $T=277$K. While only a $5\%$ increase of the water
surface tension (SPC/E and real water) is estimated from available
data~\cite{alejandre} the viscosity depends strongly on temperature:
the relative increase has been reported to be between $55-75\%$ for
SPC/E water ($85\%$ for real water).\cite{vangunsteren} Inspecting the
MD data and considering the surface tension increase we find indeed a
large decrease in viscosity of 65$\%$ with a best-fit
$\eta_{\infty}=8.48\cdot 10^{-4}$Pa$\cdot$s.  Both systems, II and
III, show that solvent viscosity has a substantial influence on bubble
dynamics as quantitatively described by our simple analytical
approach.  In systems IV and V we return to $T=300$K but increase the
pressure $P$ by a factor of 2000 and 1000, respectively. Best fits
provide viscosities which are around 10$\%$ smaller than at normal
conditions in agreement with the very weak pressure dependence of the
viscosity found in experiments~\cite{bett,sengers} at T=300K. The
major contribution to the faster dynamics comes explicitly from the
pressure terms in eq.~(\ref{RP3}).  Although moving away from
liquid-vapor coexistence by increasing the pressure up to 2000bar we
assume (and verify hereby) that the bubble interface tension can still
be described by $\gamma_\infty$.

In system VI we investigate the influence of a hydrophilic solute on
the bubble interface motion in order to make connection to cavitation
close to molecular (protein) surfaces. As a simple measure we fix a
divalent ion at the center of the bubble so that we retain spherical
symmetry. The ion is modeled by a Lennard-Jones (LJ) potential $U_{\rm
  LJ}(r)=4\epsilon[(\sigma/r)^{12}-(\sigma/r)^{6}]$ with $Q=+2e$ point
charges and uses the LJ parameters of the SPC/E oxygen-oxygen
interaction. As demonstrated recently the LY equation can be modified
to include dispersion and electrostatic solute-solvent interactions
explicitly,\cite{dzubiella:prl} which extends (\ref{RP3}) to
\begin{eqnarray}
\label{RP4}
\dot R & = & -\frac{R}{4\eta_{\infty}}\left(1+\frac{\delta_{\rm vis}}{R}\right)\\
& \times &\left(\Delta P +\frac{2\gamma_{\infty}}{R}\left[1-\frac{\delta_{\rm T}}{R}\right]
-\rho_0 U_{LJ}(R) +\frac{Q^2}{32\pi\epsilon_0R^4}\right).\nonumber
\end{eqnarray}
The last term in (\ref{RP4}) is the Born electrostatic energy density
of a central charge $Q$ in a spherical cavity with radius $R$ with low
dielectric vapor $\epsilon_{\rm v}=1$ surrounded by a high dielectric
liquid ($1/\epsilon_{\rm l}\simeq 0$). The electric field around the
ionic charge and the dispersion attracts the surrounding dipolar water
what accelerates and eventually completely governs the bubble collapse
below a radius $R(t)\lesssim 13$\AA~ ($t\gtrsim 7$ps) as also shown in
Fig.~2. The theoretical prediction (\ref{RP4}) agrees very well
without any fitting using the viscosity from system I. We find that
the acceleration is mainly due to the electrostatic attraction; the
dispersion term plays just a minor role while the excluded volume
repulsion eventually determines the final (equilibrium) radius of the
interface with $R(t=\infty)\simeq 2$\AA.

\section{Conclusions}

In conclusion, we have presented a simple analytical and quantitative
description of the interface motion of a microscopic cavity by
modifying the macroscopic RP equation. Based on our MD data we find
for the macroscopic description that analogous to the surface tension
the viscosity has to be corrected for curvature effects, a prediction
compelling to investigate further in detail and probably related to
the restructuring of interfacial water for high curvatures (small
$R$). The viscosity correction accelerates collapse dynamics markedly
below the equilibrium crossover scale ($\sim$1nm) in contrast to the
pure equilibrium picture where surface tension decreases what slows
down the collapse.  Further, we find that the dynamics is
curvature-driven due to the {\it corrections} to surface tension and
viscosity, not due surface tension as often postulated.\cite{spohn} As
a simple estimate, the interface velocity is typically given by the
ratio of surface tension and fluid viscosity, $v\simeq \gamma/\eta$.

A comment has to be made regarding the recent work of Lugli and
Zerbetto on MD simulations of nanobubble collapse in ionic solutions.
While their MD data of the collapse velocity for a 1nm bubble agree
very well with our results their interpretation in terms of the RP
equation is different. They fit the 'violent regime' solution of the
RP equation to the data [which is the solution of only the inertial
part, left hand side of (\ref{RP1})] and argue that the violent regime
still holds on the nm scale. As demonstrated in this work, we arrive
to a different conclusion: the collapse is friction dominated, the
collapse driving force is mainly capillary pressure, and we suggest
that the microscopic viscosity has to be curvature corrected to
explain the high curvature collapse behavior in the MD simulations.
The good agreement between our modified RP equation and the MD data
for different solvent conditions, leading for instance to an altered
solvent surface tension or viscosity, support our view.

We finally note that extensions of the LY equation are based on
minimizing an appropriate free energy $G(R)$ or free energy functional
\cite{boruvka,dzubiella:prl} so that we can write in a more general
form $\dot R \sim [\partial G(R)/\partial R]/[\eta(R)R]$.  It is
highly desirable to generalize this simple dynamics further to
arbitrary geometries with which a wide field of potential applications
might open up, i.e. an efficient implicit modeling of the water
interface dynamics in the nonequilibrium process of hydrophobic
nanoassembly, protein docking and folding, and nanofluidics.

\section*{Acknowledgements}

J. D. thanks Lyderic Bocquet for pointing to the RP equation, Bo Li
(Applied Math, UCSD), Roland R. Netz, Rudi Podgornik, and Dominik
Horinek for stimulating discussions, and the Deutsche
Forschungsgemeinschaft (DFG) for support within the
Emmy-Noether-Programme.

\end{document}